# Role of Random Interaction Connection in the Order Transition of Active Matter Based on the Vicsek Model

Ruizhi Jin and Kejun Dong*

*School of Engineering, Design and Built Environment, Western Sydney University, Sydney, Australia*

*Corresponding author: Kejun.Dong@westernsydney.edu.au

**ABSTRACT**. Randomness plays a key role in the order transition of active matter but has not yet been explicitly considered in pairwise interaction connection. In this *letter*, we introduce the perception rate $P$ into the Vicsek model as the probability of the interaction connections and model the connections as superposition states. We show that with increasing $P$, the polar order number undergoes an order transition and then saturation. The order transition is a first-order phase transition with band formation, and the effect of $P$ is different from density. The change of the order number is linked with the interaction structure. The order transition, order saturation, and phase separation correspond to different critical changes in the local interaction number. The global interaction structure is further analyzed as a network. The decrease of $P$ is comparable to random edge removal, under which the network experiences modal transitions near the critical points of the order number, and the network exhibits surprising robustness. Our results suggest that random interaction can be a new important factor in active matter models, with potential applications in robotic swarms and social activities.

## I. INTRODUCTION

It has long been observed that in active matter, collective ordered behaviors emerge from the seemingly disordered actions of individuals, which have fascinated researchers across disciplines [1-3]. One kind of the ordered behaviors is polar flock, i.e., a collect of random mobile individuals can move along a common direction, which is seen across a wide range of scales from mammal herds and bird flocks [4] to bacteria colonies [5] and molecular motor [6]. Many studies suggest that simple local interaction rules are sufficient to generate this order behavior [7], which could be general laws across different scales and species [8,9].

The self-propelled particle (SPP) model is often used as a basic model to study active matter. In 1995, Vicsek et al. [10] proposed a model in which particles are driven with a constant absolute velocity but their velocity directions are aligned with their neighbors with random perturbations. This model displays the polar order phase transition when the perturbation strength is decreased [10,11]. Then a continuum method based on modified hydrodynamic equations was developed to describe the macroscopic collective behavior of the Vicsek system [12], which revealed a unique long-range ordered state [13]. Since it is difficult to solve the hydrodynamic equations explicitly, the mean-field approaches were applied to simplify those equations, which can still capture the essential features of the Vicsek model [14,15]. Studies based on the Vicsek model and its continuous form have revealed that the polar order transition depends on various controlling variables and the Vicsek system exhibits a wide variety of features, such as continuous and discontinuous phase transitions [16], phase separation [17], and the formation of a giant cluster without cohesion [18].

The collective behavior of a system essentially results from the interaction between individual members. The great potential of the Vicsek model encouraged researchers to explore different interaction rules to study different systems, such as topological interaction [7], limited range of perception cones [19], and additional repulsion and attraction [3,20]. The topological interaction is better than the metric interaction in reproducing the collective behavior of starling groups [7], whereas diverse [21] and heterogeneous interactions [20,22,23] have also been considered in other studies.

However, an important aspect of the interaction that has not been explicitly considered is the randomness of the interaction connection. Randomness plays a key role in order transition [24]. It is normally modeled at the level of individuals such as the noise in the velocity direction. Recent studies also considered the randomness of the interaction strength [25]. However, the selection of interaction pairs is deterministic, based on either metric [10] or topological [7] distance in previous models. In small-world networks such as neural networks, the connection between individuals may be connected randomly in space [26,27], and the connections may randomly appear or disappear over time [28]. This randomness factor can be added to SSP models to capture uncertainty in biotic and social interactions in a new perspective.

To consider the randomness in the interaction connection, in this letter, we introduce a new parameter, the perception rate ($P$), into the Vicsek model to quantify the probability of two particles having interaction. We numerically study the collect behavior of the system with the change of the perception rate and characterize the related changes in the interaction structure. As $P$ increases, the system exhibits a polar order transition followed by order saturation, accompanied by interesting dynamics and structural changes. Moreover, the order transition and saturation can be related to the modal transitions of the system's network structure. The study shows that the interaction randomness can be a new important factor in active matter models, which provides alternative and complementary mechanisms to other interaction models and shows potential applications in robotic swarms and social activities [29].

## II. VICSEK MODEL WITH PERCEPTION RATE

The Vicsek model with the perception rate is based on the original Vicsek model [10], which describes $N$ point-like particles moving with constant velocity magnitude $V$ in a two-dimensional box of $L \times L$ with periodic boundary conditions. The positions and velocities of particles are updated when the system evolves for one time step $\Delta t$. At each time step, the velocity direction of each particle is updated by the average velocity direction of all particles (including itself) within the distance of $r_0$, plus a noise randomly generated between $[-\eta/2, \eta/2]$, where $\eta$ is the noise strength, given by:

$$\theta_i(t+\Delta t) = \text{atan2}\left(v_{i,y} + \sum_{j=1}^{N_b} v_{j,y}, v_{i,x} + \sum_{j=1}^{N_b} v_{j,x}\right) + \varepsilon \quad (1)$$

where $\theta_i$ is the velocity direction of particle $i$, $\varepsilon$ is the noise, $v_{i,x}$ and $v_{i,y}$ are the x and y components of the velocity of particle $i$ respectively, $v_{j,x}$ and $v_{j,y}$ are the $x$ and $y$ components of the velocity of the $j$th neighboring particle, and $N_b$ is the number of the neighbors of particle $i$. The particle position at the next time step is calculated based on the new velocity, which is the forward-difference method mostly used in current Vicsek-like models [8].

To consider the randomness of the interaction connection, we assume particles have a perception rate of $P$ ($0 \leqslant P \leqslant 1$). At each time step, for each pair of neighboring particles $i$ and $j$ identified by their distance, a random number $q$ uniformly distributed in [0, 1) is generated, and if the number is smaller than $P$, particles $i$ and $j$ will consider each other in their velocity alignment, otherwise they will be invisible to each other. Correspondingly, Eq (1) is changed to:

$$\theta_i(t+\Delta t) = \text{atan2}\left(v_{i,y} + \sum_{j=1}^{N_b} H_{ij} v_{j,y}, v_{i,x} + \sum_{j=1}^{N_b} H_{ij} v_{j,x}\right) + \varepsilon,$$

$$H_{ij}(q) = \begin{cases} 1 & q \leqslant P \\ 0 & q > P \end{cases} \quad (2)$$

Note $q$ will be randomly generated for each pair of neighboring particles at each time step, and $H_{ij} = H_{ji}$ i.e., particles $i$ and $j$ will be either mutually perceptible or mutually invisible. The value of $P$ controls the likelihood of interaction connection. When $P=1$, all neighboring particles interact with the central particle, which is the same as the original Vicsek model. When $P<1$, interaction is always based on probability, as shown in FIG 1.

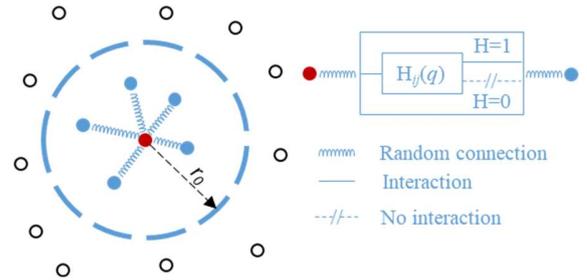

FIG 1. Schematic diagram of random connections between a central particle and its neighboring particles in the modified Vicsek model.

The adaptation to the Vicsek model here differs from the random interaction strength in the previous study [25]. By using the step function $H_{ij}$ as a weighting parameter, the interaction between a pair of particles can be regarded as a superposition of "interaction" and "non-interaction" states, which may give the interaction a quantum-like behavior. On the other hand, the resulting random connections are similar to the Erdős–Rényi network model [30], where a given number of edges connecting nodes are randomly assigned. However, here the possible connections are between neighboring particles and each connection is determined based on probability, while the connections are not randomly assigned, and the total connection number is not fixed.

The order number of the system, denoted as $\Phi$, is calculated as $\Phi = \frac{1}{NV}|\sum_{i=1}^{N} \mathbf{V}_i|$, where $\mathbf{V}_i$ is the velocity of particle $i$. We choose a set of simulation parameters used in the original Vicsek work [10], which are listed in Table S1 [31]. These parameters yield an ordered number above 0.9 when $P$ is 1 so that we can study the increase of $\Phi$ with increasing $P$. The simulation procedure is also the same as the Vicsek model. At $t=0$, we randomly assign the initial position and moving direction to each particle and $\Phi \approx 0$. Then the system iterates, and the order number is monitored. We use the time-averaged data in the period when the system reaches a steady state (see FIG S2 [31]).

## III. RESULTS
### A. Order transition with $P$

FIG 2a shows how the order number $\Phi$ changes with the perception rate $P$ under different densities. There are generally two critical points. When $P$ increases from 0 to the first critical point, $\Phi$ stays nearly zero; but once $P$ surpasses this point, $\Phi$ presents a sharp increase, indicating a clear order transition. Thus, this first critical point is defined as the order transition point and denoted as $P_c$. Then $\Phi$ keeps increasing until $P$ reaches the second critical point, after which $\Phi$ remains almost unchanged until $P=1$. Therefore, we define this critical point as the saturation point, denoted as $P_s$. The saturation phenomenon is also observed in the original Vicsek model [32]. The values of $P_c$ and $P_s$ are identified by the Binder cumulant number and derivative $d\Phi/dP$ respectively (see FIG S3[31]). Under different densities, $\Phi$ always shows similar changes with $P$, and generally $P_c$ is smaller than 0.01 while $P_s$ is smaller than 0.5. However, quantitatively $\Phi$ as well as $P_c$ and $P_s$ are dependent on $\rho$ (detailed values in Table S2 [31]). With increasing $\rho$, both $P_c$ and $P_s$ shift to a lower $P$, and the increase of $\Phi$ after $P_c$ is steeper.

The nature of phase transitions in Vicsek-like systems remains a topic of significant interest [24]. The order transition induced by increasing $\rho$ or decreasing $\eta$ in the original Vicsek model [10] was found to be a second-order phase transition. However, later studies showed it turns into a first-order phase transition when $L$ [16] or $V$ [33] is over a critical value, or the noise type is changed to vectorial noise [34]. Moreover, the band formation at the onset of the order transition is a signature of a first-order phase transition [5], which has been explained by density-dependent polar field fluctuations [24,35] and anisotropic diffusion [33].

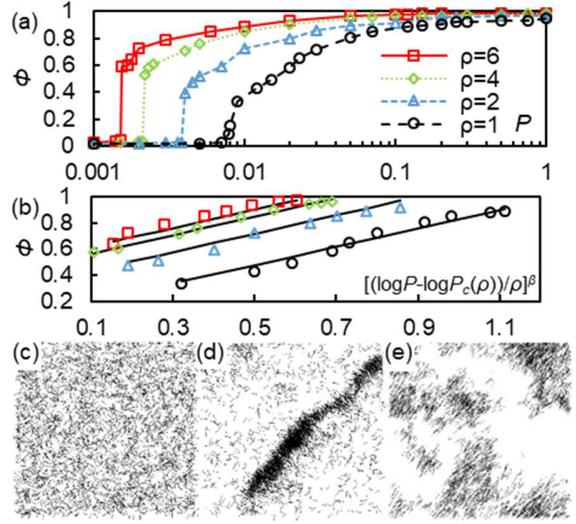

FIG 2. (a) Polar order number $\Phi$ versus perception rate $P$. (b) $\Phi$ versus $[(\log P - \log P_c(\rho))/\rho]^\beta$ at various densities (see legend in (a)); four predictive lines have the same slope of 0.7. (c)-(e) are simulation snapshots with different $P$ at $\rho = 4$, where (c) $P = 0.001$, $\Phi = 0.02$, (d) $P = 0.005$, $\Phi = 0.76$, (e) $P = 0.5$, $\Phi = 0.98$.

Interestingly, although our system has a similar size and velocity to the original Vicsek model showing a second-order phase transition, here the order transition at $P_c$ is a first-order phase transition (see FIG S3 [31]). In addition, FIG 2c-e illustrate that as the system transitions from a homogeneous disordered state to a homogeneous ordered state, a traveling band is formed when $P$ is slightly higher than $P_c$ but disappears as $P$ increases further, which is similar to the phase separation associated with the first-order phase transition found in other studies [17]. The observation further indicates that the perception rate brings new effects on the order transition of Vicsek-like systems, particularly on the formation of traveling bands [5], which deserves further studies.

In the original Vicsek model, the increase of the order number is correlated to the controlling variables in a similar form to the phase transition of an equilibrium system, e.g., $\Phi \sim [(\eta_c(\rho)-\eta)/\eta_c(\rho)]^\beta$ [10]. Here we find that between $P_c$ and $P_s$, $\Phi$ can be correlated to $P$ by $\Phi \sim [(\log P - \log P_c(\rho))/\rho]^\beta$ with a constant ratio of 0.7 independent of density, as shown in FIG 2b. $\beta$ obtained by data fitting is 0.42, which agrees with previous studies [36]. The universal relationship suggests general mechanisms, which are studied by analyzing the interaction structure of the system.

## B. Interaction number

Some important studies suggest topological interaction instead of metric interaction in Vicsek systems, i.e., a particle interacts with a fixed number of neighbors, identified by the metric distance [7] or by Voronoi method [37]. We define the number of neighbors included in the velocity alignment as the interaction number, denoted as $K$. Since changing $P$ leads to a change in $K$, in light of the order transition with topological interaction [38], we investigate the relationship between the order number and $K$.

For the four series of cases with different densities shown in FIG 2a, we plot the normalized order number, $\Phi/\Phi_m$, as a function of the average interaction number $\langle K \rangle$ in FIG 3a, where $\Phi_m$ is the value of $\Phi$ at $P=1$. Compared to FIG 2a, the data of different densities show a more uniform correlation between $\langle K \rangle$ and $\Phi/\Phi_m$, especially at the two critical points. Interestingly, $\Phi$ increases with the increase of $\langle K \rangle$ and reaches saturation when $\langle K \rangle$ is above 6, which agrees with the previous topological interaction models that increasing topological interaction number from 1 to 3 significantly improves the order number [7,38], and a highly ordered state is achieved when the topological interaction number is 6 [7,37]. However, unlike those models, here $K$ is not fixed and varies in a wide range.

Similar to previous studies on the topological interaction number [38], we study how particles with different $K$ affect the order number by calculating their angular deviation from the velocity direction of the system, given as, $\delta = \arccos(\frac{\mathbf{v}_i \cdot \mathbf{v}_a}{|\mathbf{v}_i||\mathbf{v}_a|})$, where $\mathbf{V}_a = \sum_{i=1}^{N} \mathbf{V}_i / N$, and $\delta$ is averaged for particles with the same $K$.

FIG 3b shows $\delta$ as a function of $K$. Generally, $\delta$ decreases with increasing $K$, as a particle with more neighbors deviates less from the group. However, the effect of $K$ varies in different ranges. $\delta$ decreases quickly when $K$ increases from 0 to 3 but the decrease becomes slow when $K$ increases from 4 to 6. When $K$ reaches 7, $\delta$ finally stops decreasing and stabilizes at a minimum value, indicating that the average moving direction of particles with $K > 6$ is consistent with the group moving direction. Notably, the change of $\delta$ with $K$ is similar under different conditions.

Based on this observation, we divide $K$ into three categories, unsaturated $K$ ($K \leqslant 3$), transition $K$ ($4 \leqslant K \leqslant 6$), and saturated $K$ ($K > 6$). We examine how the fraction of each category changes with $P$. As shown in FIG 3c, when $P < P_c$, the faction of unsaturated $K$ is 100%, and $\Phi \approx 0$. When $P$ just exceeds $P_c$, the fraction of unsaturated $K$ begins to decrease while that of transition $K$ begins to increase, leading to a sharp increase in $\Phi$ (FIG S4 [31] shows a closer look). The fraction of saturated $K$ remains 0 but begins to increase when $P$ surpasses 0.02. This value is between $P_c$ and $P_s$, while interestingly it coincides with the disappearance of the traveling band (see animations in [31]), indicating that the appearance of the particles with saturated $K$ may initiate ordered clusters and hence destroy the band. The saturated $K$ becomes the major category when $P$ just exceeds $P_s$, corresponding to the saturation of the order number. The changes are similar under other densities as can be seen from FIG S5 [31].

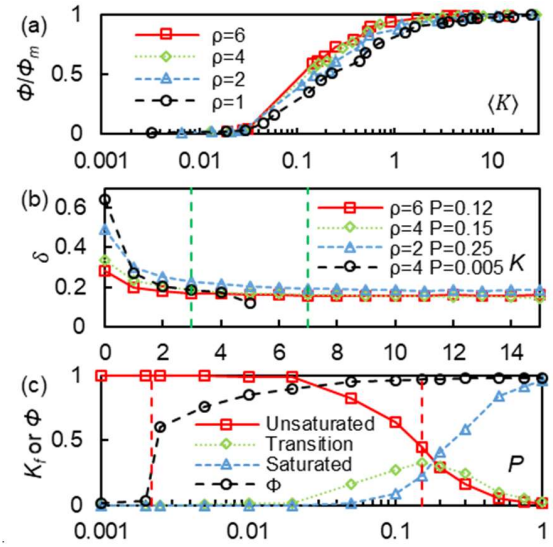

FIG 3. (a) Normalized order number $\Phi/\Phi_m$ versus average interaction number $\langle K \rangle$ at various densities. (b) Angular deviation $\delta$ versus $K$ under different conditions. Two green dash lines divide $K$ into Unsaturated $K$ (0-3), Transition $K$ (4-6), and Saturated $K$ (>6). (c) Fractions $K_f$ of three $K$ categories and $\Phi$ versus $P$ at $\rho=4$. Two red dash lines mark $P_c$ and $P_s$ respectively.

Unlike topological interaction, our model does not fix the interaction number. However, the critical interaction number related to the onset of order transition agrees with those found in the topological interaction models. In addition, we observe that band formation and disappearance are also associated with the critical changes of $K$. Therefore, our model can complement the topological interaction models.

### C. Comparison with $\rho$

In the original Vicsek model, the polar order transition can occur when the density increases beyond a critical value [10], which is also related to the interaction number. In the mean-field approach, the average number of interacting particle neighbors is assumed to be proportional to the density, $\langle K \rangle = \pi r_0^2 \rho$, which is linked with the order transition [32]. It could be thought that varying $P$ and varying $\rho$ have similar effects on the mean interaction number and then the order number. However, as shown above, macroscopically the order number is related to $1/\rho \cdot \log P$ in our system, whereas in the original Vicsek model the order number is related to $\rho$. To understand the difference, we compare the effects of $P$ and $\rho$ on $\Phi$ and $\langle K \rangle$ simultaneously. We define the equivalent density in our system as $\rho_{eff} = P \cdot \rho$ and simulate two series with $\rho_{eff}$ changing from 0.001 to 4. In the first series, $P$ is fixed at 1, and $\rho$ changes from 0.001 to 4, and we define this series as the $\rho$ series. In the second series, $\rho$ is fixed at 4, and $P$ changes from 0.001 to 1, and we define this series as the $P$ series.

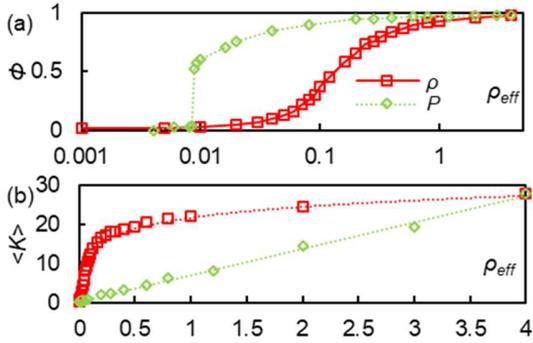

FIG 4. (a) Polar order number, and (b) average interaction number, versus effective density. The $\rho$ series in red are obtained by the original Vicsek model ($P=1$) under different $\rho$, while the $P$ series in green are obtained by varying $P$ with $\rho$ fixed at 4. The red dotted line in (b) represents a logarithmic fit applied to the $\rho$ series and the green dotted line represents a linear fit applied to the $P$ series.

FIG 4 compares the two series. Interestingly, although increasing both $P$ and $\rho$ can increase $\Phi$ and $\langle K \rangle$, the quantitative relationships are obviously different. Firstly, FIG 4a shows that the order transition induced by $P$ is a first-order phase transition while that induced by $\rho$ is a second-order phase transition. FIG 4b shows that $\langle K \rangle$ is in a linear relationship with $P$ but in a complicated relationship with $\rho$, which may be responsible for the different types phase transitions. In addition, under the same $\rho_{eff}$, $\langle K \rangle$ is smaller in the $P$ series than in the $\rho$ series, but $\Phi$ is higher. This indicates that by introducing $P$, the system can achieve a higher order number with fewer interactions than the original Vicsek model, thus showing a higher alignment efficiency. It is probably because in the original Vicsek model, when the density is small, clusters of high-density and high-order are formed locally [34] due to the local coupling between order and density [39,40]. These segmented clusters break the homogenous state assumed in the zero velocity limit [41] or mean-field approximation [15], which complicates the relationship between $\langle K \rangle$ and $\rho$. In addition, the clusters have high interaction numbers but may decrease the global order numbers as their velocity directions are different [10]. On the contrary, in our model particles are more uniformly distributed, which is beneficial for information spread in the system [27] and improves the global order.

### D. Network analysis

The above analysis indicates that the order transition depends not only on neighbors but also on the interaction structure of the whole system. Recently, network models have been employed to characterize the structures of complex dynamics systems, such as the molecular interaction network in the supercritical fluid system [42] and the protein interaction network in the chemotaxis system [43]. They have also been used to analyze the structure of the Vicsek systems [5,44] and to derive phase transition points [45].

Chaté suggested the criterion for applying the network model to the Vicsek system is that $L$ should be smaller than the crossover size to avoid strong dynamics effects [39]. In our system, $L$ and $V$ are the same as the original Vicsek model [10]. Therefore, similar to previous studies [5,44], we analogize the system to a complex network where each particle represents a node, and two nodes are connected with an undirected edge if they have alignment interaction with each other. It is found that the edge number in the system decreases linearly with $P$ (see FIG S6 [31]), thus decreasing $P$ from 1 to 0 causes a similar effect analogous to the damage to the network due to random edge removal [46]. Similar to the network research [46], we study how the reduction of edges affects the network structure of the system and its functionality, i.e., the order number.

We use two parameters to characterize the network structure, namely, the probability density distribution of node degree, $n(K)$, and the dimensionless size of the largest component, $N_c$, which is the number of the particles in the largest component divided by the total particle number, and here a component refers to a set of particle nodes connected together. Note both n($K$) and $N_c$ are time-averaged values in the steady state [31]. The form of n($K$) is commonly used to characterize a network model [47]. Interestingly, n($K$) in our system presents three typical distributions at different stages.

In FIG 5, we show the modal transitions of the network in three stages when $\rho$=4, whereas under other densities the results are similar (see FIG S8 [31]). FIG 5 shows when $P$=1, almost all particles are connected in the largest component as $N_c \approx 1$. In stage I (FIG 5a), as $P$ decreases from 1 to $P_s$ (0.15), $N_c$ decreases slightly and n($K$) always follows a normal distribution $\frac{1}{\sqrt{2\pi}\cdot\sigma}e^{-\frac{(K-\langle K \rangle)^2}{2\sigma^2}}$ (FIG 5b) with $\sigma \approx \langle K \rangle/2$ (Table S3 [31]), which is also similar to the distribution of $K$ in the clusters in other Vicsek systems [44] (FIG S7 [31]). This indicates that the network is very robust and maintains similar structures. Since this stage coincides with the saturation stage of $\Phi$, the consistency indicates that the structure robustness supports the functional robustness, and the network displays a high degree of tolerance against the random edge removal, similar to a scale-free network [46].

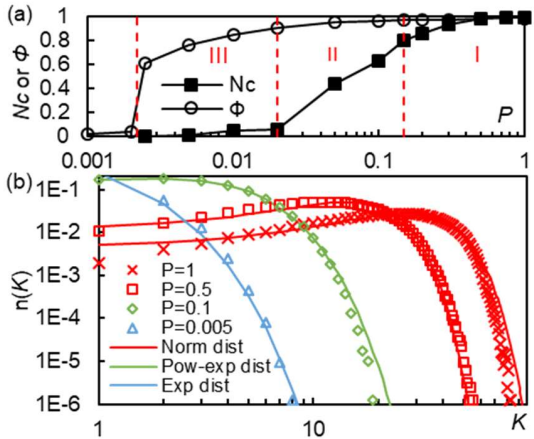

FIG 5. Complex network analysis of the system when $\rho$=4. (a) Changes in network topology connectivity and functionality as $P$ changes. Topology connectivity is represented by the dimensionless largest component size and the functionality is represented by the order number. (b) The different distribution types of n($K$) at different stages marked in (a).

In stage II, as $P$ decreases from $P_s$ (0.15) to 0.02 (below which the traveling band appears), the network structure is critically different from stage I: the topological breakdown of the full network occurs as $N_c$ drops significantly (FIG 5a), and n($K$) changes to power-law distribution with an exponential decay, $e^{-B(K+\langle K \rangle)}(K+\langle K \rangle)^C$ (FIG 5b). Despite the network edges drop below 15% (as $E$ is proportional to $P$), $\Phi$ decreases slightly from 0.97 to 0.9. The maintenance of the high order number may be due to two reasons. First, a giant component containing more than 50% nodes ($N_c$>0.5) remains in some cases, which can keep the system's function stable as shown in other networks [48]. Secondly, n($K$) distribution is similar to those of the small-world models [49,50] and social networks such as Orkut and Weibo [51], which can have positive long-term effects on the collective action of the system. This suggests a similarity between the studied system and those social networks in terms of the relationship between structure and function.

In stage III, $P$ decreases from 0.02 to $P_c$ (0.0022). The network changes to a single-scale network as n($K$) becomes a simple exponential decay function $e^{-BK}$ (Table S3 [31]), which is mainly seen in some economic networks with strong constraints [50]. $N_c$ is below 0.1, indicating only small components exist. Moreover, previous analysis shows that in this stage, the saturated $K$ vanishes, and the traveling band appears (see animations in [31]). Although simultaneously $\Phi$ drops to 0.5, much smaller than the saturation stage, it is still relatively high, probably because of the traveling band. Though the band contains only small components, it broadcasts the alignment interaction to a large number of isolated nodes when sweeping across the whole region. This indicates the conducive effect of dynamics on maintaining the network function at low connections. When $P < P_c$, $E$ and $\Phi$ both decrease to near zero, indicating the network is completely disintegrated.

Although our system is not fully equivalent to a network, the above analyses indicate the interaction structure plays an important role in the critical changes of the order number, and the collective behavior can be related to network research. On the other hand, the network behaviors of our system inspire some favorable effects of randomness and dynamics on the network robustness. It is also worth comparing with the Vicsek models with topological interaction [38], which are expected to generate a regular network with most nodes having the same degree. Such a regular network may not capture the features of biological systems [27], while our system exhibiting small-world network structures may be closer to those systems.

## IV. CONCLUSION

In summary, we have studied the effect of interaction randomness on the order transition of active matter by introducing the perception rate $P$ in the alignment interaction in the Vicsek model. Through numerical simulations, we find that as $P$ increases from 0 to the order transition point $P_c$ (a small value), the system undergoes a first-order phase transition accompanied by band formation. Beyond $P_c$, the order number increases sharply until $P$ reaches the saturation point $P_s$, which is far below full perception ($P=1$). Beyond the saturation point, the order number remains nearly constant. Between the two critical points, the effect of $\log P/\rho$ on $\Phi$ is similar to that of noise or density in the previous Vicsek model.

It is found that the critical change of the order number is closely related to the interaction structure. The order transition at $P_c$ nearly coincides with the emergence of particles with interaction number $K \geq 4$, the traveling band can be found before the emergence of particles with $K > 6$, and the saturation of the order number at $P_s$ coincides with the average interaction number $\langle K \rangle$ exceeding 6. The critical interaction number agrees with those found in topological interaction models, though our model does not fix the interaction number, and the critical values are naturally developed, suggesting the interaction randomness can complement those models. The global interaction structure is further analyzed by network models. The decrease of $P$ is analogous to the random edge removal damage of the network, under which the network undergoes modal transitions at the critical points. Interestingly, the network of our system displays a surprising level of robustness in response to a large reduction in edge number, especially for its function, i.e., the order number.

Our finding shows that interaction randomness can be a new and important controlling variable for the collective behavior of active matter, such as in controlling the phase transition type and phase separation. Our analysis reveals connections between the order transition and the interaction structure, which is similar to the recent study on the network model [52] and the previous study on the order transition of granular particles [53], providing a new perspective on the order transition in active matter.

Our results also provide some insights into applications. For example, introducing certain randomness in the communication of robot swarms may not lose stability but improve efficiency, and a specific distribution of connection will be resilient to communication interference. The similarity between our system's network model and social networks also encourages the application of our model to those systems, especially for understanding how interaction randomness influences complex social behaviors such as crowd control [20,54] and voting [55].

## ACKNOWLEDGMENTS

The authors are grateful for the financial support from the Australian Research Council (IH230100010) and international PhD scholarships from Western Sydney University.